# Broadband Self-Switching of Femtosecond Pulses in Highly Nonlinear High Index Contrast Dual-Core Fibre

M. Longobucco[1,2], I. Astrauskas[3], A. Pugžlys[3,4], D. Pysz[1], F. Uherek[5], A. Baltuška[3,4], R. Buczyński[1,2], I. Bugár[1,5]

[1]*Department of Glass, Łukasiewicz Research Network - Institute of Electronic Materials Technology, Wólczyńska 133, 01-919 Warsaw, Poland*
[2]*Department of Geophysics, Faculty of Physics, University of Warsaw, Pasteura 5, 02-093 Warsaw, Poland*
[3]*Photonics Institute, Vienna University of Technology, Gusshausstrasse 27-387, 1040 Vienna, Austria*
[4]*Center for Physical Sciences & Technology, Savanoriu Ave. 231, LT-02300 Vilnius, Lithuania*
[5]*International Laser Centre, Ilkovičova 3, 841 04 Bratislava, Slovakia*

In this paper, we demonstrate a pulse energy-governed nonlinear self-switching of femtosecond pulses at wavelength of 1700 nm in highly nonlinear high index contrast dual-core fibre. The fibre structure is composed of two microdimension cores and a solid cladding made of two thermally matched soft glasses developed in-house. Double switching behaviour under monotonic increase of the input pulse energy from 1 nJ to 3 nJ was observed at 35 mm fibre length in its anomalous dispersion region. A spatial intensity distribution measurement identified a switching contrast of 16.7 dB, and the registered spectra of the two output channels revealed a broadband switching behaviour within 150 nm. The obtained results indicate a novel type of solitonic switching and represent a significant progress in comparison to previous experimental works and has a high application potential.

## 1. Introduction

The high-speed, optical fibre based long distance data transport demands for compatible devices for switching [1], routing [2] or buffering [3] the signals in an optical way. To facilitate this challenge, theoretical studies have been intensively investigating the nonlinear directional coupler approach by using dual-core fibres, which promises simple and compact all-optical switching [4–6]. This dual-core fibre based switching technique has many advantages over other approaches. For instance, ultrashort pulses in rather-short dual-core fibres (DCF) can be self-switched while the nonlinear optical loop mirror requires meter-long fibres to induce sufficient nonlinear phase shift and, in the case of silica fibres, tens of nanojoule pulse energies [7]. Another example is the fibre grating coupler using chalcogenide fibres [8,9]. Its disadvantage over the dual-core approach resides in its need for a spectrally shifted pump and signal pulses and wavelength division multiplexers to combine them at the input and separate at the output. Finally, such type of period grating exceeds the length of 100 mm, which is significantly longer than the optimal fibre lengths in the case of DCFs. Several works demonstrated self-switching performance of dual-core fibres already around 10 mm fibre length at potential transfer rates beyond 1 THz/s [1,10,11]. The first experiment of ultrafast switching using dual-core fibre has been performed in the normal dispersion region and revealed severe pulse distortions [10]. Therefore, the attention of nonlinear fibre community was drawn towards solitonic switching, which promises nonlinear control of the ultrafast pulses with simultaneous maintenance of their amplitudes and phases [4,12]. Recently a theoretical book chapter was published, which review many aspects of the solitonic dual-core propagation [13]. It comprises among others the asymmetry and polarization effects, problems of dual-core fibre lasers and parity-time symmetry. On the experimental side, we have demonstrated for the first time a true switching of femtosecond pulses in solitonic regime using a soft glass dual-core photonic crystal fibre (PCF) at the excitation wavelength of 1650 nm [11]. The experimental results were supported by corresponding numerical simulations confirming the mutual effect of high order soliton fission [14] and wavelength dependent coupling oscillations. The best switching has been achieved at 14 mm fibre length and tens of nanojoule pulse energies, however the high switching contrasts were demonstrated only in limited segments of the nonlinearly broadened spectra.

In order to improve the switching performance further, we chose a new fibre material with 20 times higher nonlinear index of refraction ($n_2$) than the standard silica or the previously used soft glass. Taking advantage of a numerical simulation tool, verified in our previous studies, we optimised an air-glass dual-core (DC) PCF design in order to support broadband switching at sub-nanojoule pulse energies [15]. The experimental demonstration in the manufactured air-glass DC PCF failed. This likely happened because of the presence of a dual-core asymmetry that impeded the DC PCF function in the envisaged controllable soliton self-trapping regime [16]. Therefore, our strategy was to devise a highly effective nonlinear switching by utilizing a novel all-solid PCF approach [17,18], which promises a higher level of dual-core symmetry. To achieve this goal, two soft glasses with a refractive index contrast of 0.4, combinable during the fibre-drawing process, have been synthetized in-house. This unusual technological approach enables broadband anomalous dispersion, which is necessary for ultrafast solitonic propagation, and also supports low-energy performance due to the highly nonlinear guiding glass utilized [19]. Using the experimentally determined optical parameters of the two glasses, we performed extensive nonlinear simulation studies in order to demonstrate effective all-optical switching in DCF. The simulations were based on the solution of Coupled Generalized Nonlinear Schrödinger Equations allowing to model the ultrafast solitonic dual-core propagation [4,12,20]. The obtained results revealed that the best structural strategy is the high refractive index contrast dual-core geometry with a simple cladding without photonic structure [21,22]. Considering this, not only the manufacturing is simplified, but the simulations also predicted higher switching contrasts (beyond 15 dB) and lower switching energies (picojoule range) than similar theoretical and experimental works.

In this paper, we present the first experimental results of solitonic all-optical switching of femtosecond pulses in high index contrast all-solid dual-core fibre (DCF). We demonstrate a novel double switching behaviour at 35 mm fibre length under monotonic increase of the input pulse energy in the frame of a thorough cut-back study. By performing separate spectral registration of both output channels, we observed broadband character of the complex switching performance, showing first switch, back switch and second switch steps. Moreover, the global switching scenario expressed switching contrast of the integral energy at the level of 17 dB, overcoming previous similar experimental works. The identified character of the nonlinear switching indicates a special solitonic propagation scenario and its possible interpretations will be discussed at the end of the paper.

## 2. Methods

### 2.1 Fibre development

The core material of the new fibre is the lead-silicate glass PBG-08. The glass has been previously used for development of nonlinear PCFs and its linear and nonlinear properties were already verified experimentally [16,23]. PBG-08 features both high nonlinear refractive index of $4.3 \cdot 10^{-19} m^2/W$ and linear refractive index around 1.9 in the near infrared (NIR) spectral region. More recently a composition of a complementary borosilicate glass was optimized to thermally match PBG-08 glass. The result was the new glass UV-710 with the refractive index around 1.5 in NIR [19]. Fig.1 presents the group index of refraction of both glasses in the visible and NIR spectral region, and their most important rheological properties. The graph confirms the high index contrast between the glasses at the level of 0.4 in the near infrared even in the case of the short pulse propagation. The values in the table reveal a very similar behaviour of the glasses under thermal treatment, therefore allowing their combination in all-solid fibre manufacturing process.

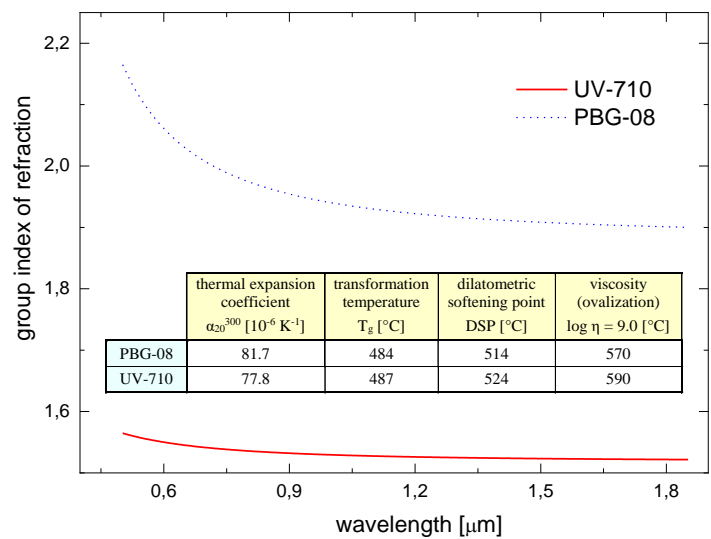

Fig.1 Group index of refraction of the two selected glasses and their key rheological properties (inserted table).

The dual-core fibre was fabricated by the stack-and-draw method, which is typically used in the case of PCF manufacturing process. After the glass synthesis, glass rods of cylindrical shape and with the same diameter were prepared from both glasses, the high index PBG-08 and the low index UV-710. The UV-710 glass rods were stacked together into a hexagonal lattice structure with 6 rings of elements around the central rod. Then, in the central line, the two UV-710 glass rods on two sides of the central rod were replaced by PBG-08 ones. Small remaining air gaps between the rods were filled during the two-step drawing process. In the first step, we manufactured a few 20 cm long subpreforms with the diameter of about 1.6 mm. In the second step, we identified the best symmetry subpreform and the final drawing process was performed. Before the final drawing process, we put the subpreform into a capillary of PBG-08 glass to form the outer cladding with larger diameter close to standard one. The final drawing resulted in a fibre with outer diameter of 111 μm. Fig.2 presents two scanning electron microscope (SEM) images of the cross-section of DCF structure with the magnification of (a) 5000 and (b) 20000. The SEM images show, that the core borders do not form the expected regular hexagons and are deformed towards a star-like shape. This deformation takes place because the UV-710 glass for the cladding is harder than the PBG-08 glass for the core area, thus the original curvature of cladding rods is preserved more. The distance between the centres of the DCF cores is 3.1 μm and effective mode area $A_{eff}$ of a single core is 1.86 μm².

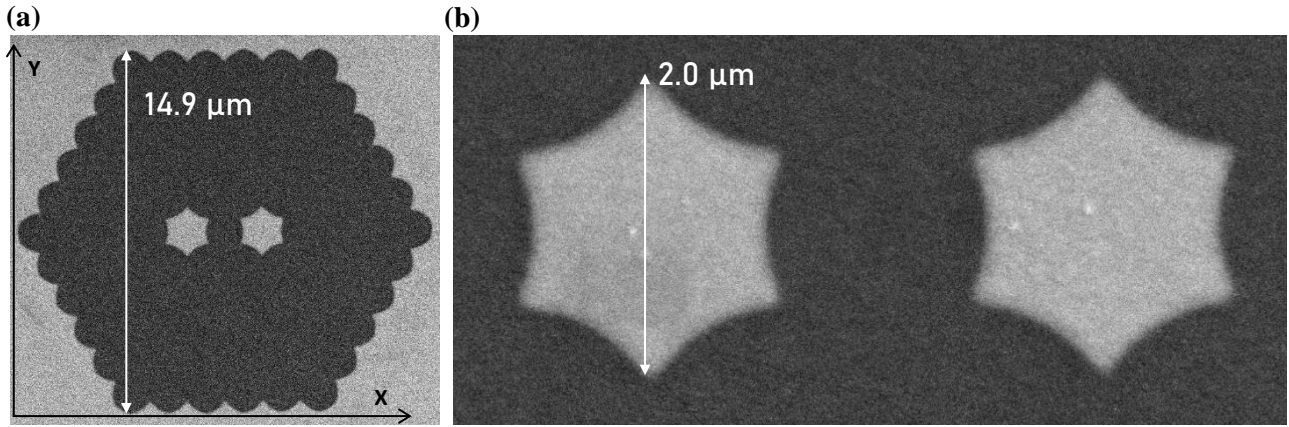

Fig.2 SEM images of the cross-section of the all-solid dual-core optical fibre structure with 6 rings around the central rod at two different magnifications: (a) 5000x, (b) 20000x.

2.2. Numerical simulation of the DCF linear optical properties

We used a commercial software (Mode Solution, Lumerical) to calculate linear optical properties of the DCF. In the frame of this work, measured material dispersion profiles of the both glasses were characterized by the Sellmeier coefficients [19]. We approximated the the DCF cross-section structure of the SEM picture by simple geometrical objects. The resulting structures are reported in Fig.3 as insets. For each core, we fitted the curvatures between the two glass borders by circles, eventually by circle sections, which are easy to apply in the mode solver. All the objects on the inset pictures that have different colours with respect to dark red represent the UV-710 glass material. At first, we analysed the single core structure in order to determine the dispersion profile. We identified a broad anomalous dispersion, which determine the character of the solitonic propagation, between the two zero dispersion wavelengths 1295 and 2360 nm, with the maximum dispersion parameter of 66 ps/km/nm at 1720 nm. The results of the simulations in Fig.3a are very similar to the case of hexagonal core structures we analysed numerically in our previous article [21]. Slight sensitivity to polarization can be observed in the graph by comparing the dispersion profiles for parallel (X) and perpendicular (Y) field oscillations with respect to the line comprising the two core centres (Fig.2a). In the second step, we analysed the dual-core structure in order to acquire the coupling length $L_c$ characteristics with the standard procedure [11,15,21]. The results in Fig.3b reveal that a stronger coupling is predicted with respect to the optimized hexagonal core structure with 3.2 μm core distance studied in [21].

In the case of the real structure, the distance between the core centres is 3.1 μm and the effective mode area of the single core mode ($A_{eff}$) slightly increased from 1.84 to 1.86 μm². Both the decreased distance and increased $A_{eff}$ enhances the coupling. This is in fact determined by the overlap integrals, which is affected by the evanescent field of one core in the area of the second core [24]. The difference between the two orthogonal polarizations of the coupling length also confirmed the sensitivity of the $L_c$ to the slight changes of the microstructure. The coupling is stronger in the case of the X polarized field, than in the case of Y polarized one, because the field oscillation vector is oriented toward the opposite core. Therefore, $L_c$ is shorter in the case of X polarized field. The numerical simulations in the case of X polarized field for the hypothetical hexagonal and real star-like core structures revealed that the coupling length at 1700 nm decreases from 9 to 5 mm, respectively. The numerical analysis of linear properties of the DCF is essential for the proper choice of the conditions of the nonlinear experiments and for the interpretation of the obtained results. It is a reliable method and provides accurate results, as it was demonstrated in the previous studies of similar DCFs [11,16]. Further reason of those calculations is, that it is not possible to evaluate experimentally the dispersion of the artificially created single core structure. The experimental determination of the coupling length at one wavelength would be possible by a thorough millimetre precision cut-back procedure. However, this method destructs the delicate fibre sample. In addition, the spectral characteristics of $L_c$ would require a tuneable laser source and permanent readjustment of both the excitation and the registration stages of the apparatus at every change of the wavelength.

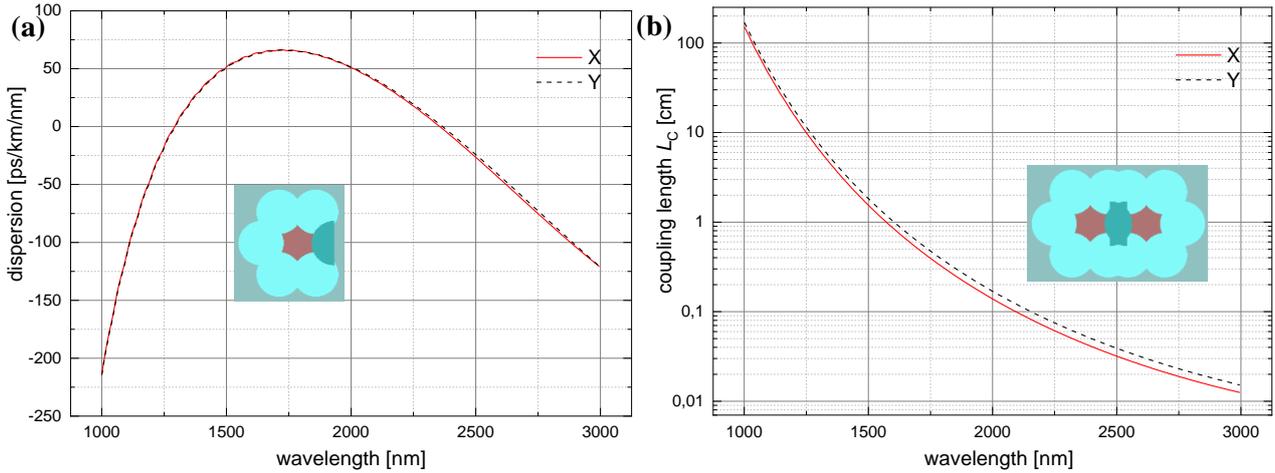

Fig.3 Spectral characteristics of the (a) single core structure dispersion and (b) dual-core structure coupling length for the both orthogonal polarization directions. The pictures of the analysed simplified structures are in insets.

2.3. Experimental investigation of the nonlinear switching performance

We performed the experimental study by using the apparatus presented in Fig.4. The source of the excitation pulses was an optical parametric amplifier (OPA) operating at 10 kHz repetition rate and tuneable in the spectral range of 1500 – 1900 nm. The two stage OPA, based on type I BBO crystals, was pumped by a frequency doubled output of a commercial Yb:KGW amplifier (Pharos, Light Conversion). Pulse duration of 70-100 fs in the whole spectral range of the idler radiation was achieved by controlling the dispersion with a prism compressor based on a pair of SF10 prisms. The energy of generated pulses exceeded a few microjoules, so we attenuated it to a few nJ level in order to prevent the damage of the DCF input facet. In order to explore the switching potential of the DCF, the pulses were directed through two half-wave plates with a Glan-Taylor polarizer placed between them. We used the first half-wave plate and the polarizer for a fine attenuation of the pulse energy, while the second half-wave plate to tune the polarization of pulses before launching them into the fibre. We managed the in-coupling and out-coupling of the radiation by two 40x microscope objectives mounted on two identical 3D positioning stages with sub-micrometre precision. The first objective enabled separate excitation of a single selected fibre core. The second objective imaged the output facet of the DCF onto a multimode collecting fibre attached to a spectrometer (NIRQuest, Ocean Optics). The default output beam path was oriented to an infrared camera chip (Xeva 1.7 320, Xenics) placing a flip mirror before the collecting fibre. The tilt of the mirror enabled the spectral registration, as it is depicted in Fig.4. We used an iris aperture before the flip mirror to restrict the spectral registration for one core only and cut out the image of the second core. The alignment of the aperture was monitored by the camera. Finally, the spectral registration was improved by focusing the light with a 25 mm objective after the aperture into the collecting fibre end.

During the experiments, we used a 1700 nm excitation wavelength, which is the longest one where the infrared camera has nearly flat spectral sensitivity. We chose the longest wavelength because the dual-core asymmetry effect is weakened by increasing wavelength, as it is shown in our previous work [16]. The DCF asymmetry lowers the power transfer ratio between the excited and non-excited core in linear regime. Therefore, it has negative effect also on the nonlinear coupling performance [24]. Preliminary experiments confirmed that, in the case of the new all-solid DCFs, the 1700 nm excitation provides better switching results than in case of the 1500 nm alternative. The advantage of the longer wavelength relies also on the dispersion profile of the fibre (Fig.3a), which reveals negligible third order dispersion (TOD) at 1700 nm. Therefore, weaker dispersive wave generation and less disturbed solitonic propagation is expected at this wavelength [14]. We optimized the pulse width empirically and we found out that the 100 fs one was more advantageous for the switching in the studied DCF in comparison to shorter pulse widths. Then, we optimized the length of the fibre by cutting it back in the range of 45 – 25 mm. We collected several series of camera images and spectra with increasing pulse energy between 0.5 – 3 nJ. In this energy range, we observed a nonlinear interaction with moderate spectral broadening. No switching above pulse energies of 3 nJ was observed. We registered the same series of camera images and spectra at every fibre length and at distinct polarization angles, in order to identify the highest switching contrasts. Dual-core fibres are inherently birefringent, therefore we analysed the effects of polarization by rotating the second half-wave plate after every cut of the fibre.

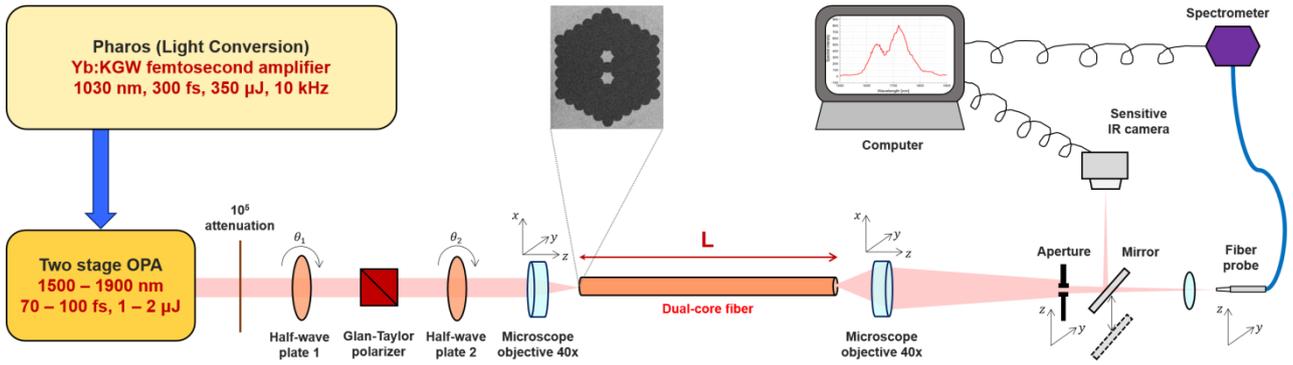

Fig.4 Experimental apparatus for the investigation of nonlinear switching in a dual-core optical fibre.

## 3. Results and discussion

In the frame of the fibre cut-back procedure, we observed the most interesting switching performance at 35 mm fibre length for X polarized field. This polarization direction is just slightly deviated from the vertical axis of the laboratory coordinates (Fig.5) as it is indicated by a white arrow on the last image taken at 2.98 nJ energy. The length of the arrow corresponds to the distance of 3.1 µm between the centres of the two cores. As seen from the camera images presented in Fig.5, a rich switching performance was observed under monotonic increase of the input pulse energy ($E_{in}$). The initial bottom core dominancy at 0.89 nJ pulse energy shifts to the top core at the energy levels 1.65 – 1.95 nJ. Further increase of the energy induces back switching with the extreme at 2.27 nJ. The second switching step to the upper core takes place between the energy levels of 2.27 and 2.62 nJ and its dominancy is maintained up to the maximum applied energy of 2.98 nJ. For every input pulse energy, we integrated the captured intensities on the images across the area of both cores individually. The obtained relative energy values $E_{bottom}(E_{in})$ and $E_{top}(E_{in})$ were used to calculate the dual-core extinction ratio as $ER = 10 \cdot \log(E_{bottom}/E_{top})$. The values of $ER(E_{in})$ are positive in the case of bottom core and negative in the case of top core dominancy, respectively. One of the most important characteristics of the switching is the switching contrast, which is the difference between the two extreme positive and negative values of the $ER(E_{in})$ series.

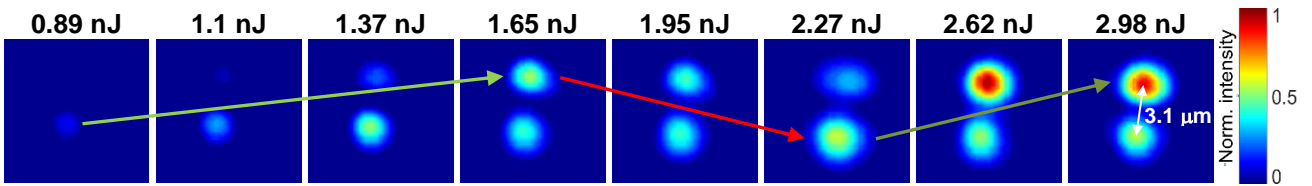

Fig.5 Infrared camera images of the output of the dual-core fibre at different energies of 1700 nm, 100 fs input pulses. The scale of the images is indicated by a white arrow on the right hand side image.

The dependence of the extinction ratio on the input pulse energy is presented in Fig.6a. As consequence of the double switching, it exhibits a change of the sign of $ER$ from positive to negative two times under monotonic increase of the pulse energy. We identified the highest switching contrast between the energy levels of 0.89 and 2.62 nJ: the switching contrast was 16.7 dB and it is marked by a vertical arrow in Fig.6a. At the same experimental conditions as the camera images were taken, we separately registered the two spectra at the outputs of each core, $S_{bottom}(\lambda)$ and $S_{top}(\lambda)$, respectively. Then, we calculated the spectrally resolved extinction ratio $ER(\lambda) = 10 \cdot \log(S_{bottom}(\lambda)/S_{top}(\lambda))$. Fig. 6b shows the dependence of the $ER(\lambda)$ on the input pulse energy at selected levels of $E_{in}$. It reveals the spectral details of the double switching behaviour. The global nonlinear switching trend is evident when increasing pulse energy from 0.71 nJ (black dotted curve) to 3.14 nJ (black solid curve). These levels approximately correspond to the minimum and maximum energies of applied pulses during the camera registration shown in Fig.5. The two corresponding $ER(\lambda)$ curves in Fig.6b are well separated between the positive and negative sections, with an exception in the area between 1720 nm and 1740 nm. Thus, they exhibit the same global switching behaviour as the corresponding camera data in Fig.6a: positive $ER$ value at low energies and negative at high energies, respectively. The inset of Fig.6b shows the spectrum for the top core in the case of the highest applied pulse energy of 3.14 nJ. It reveals that, when this maximum pulse energy is applied, the switching covers the broadest registered spectral bandwidth.

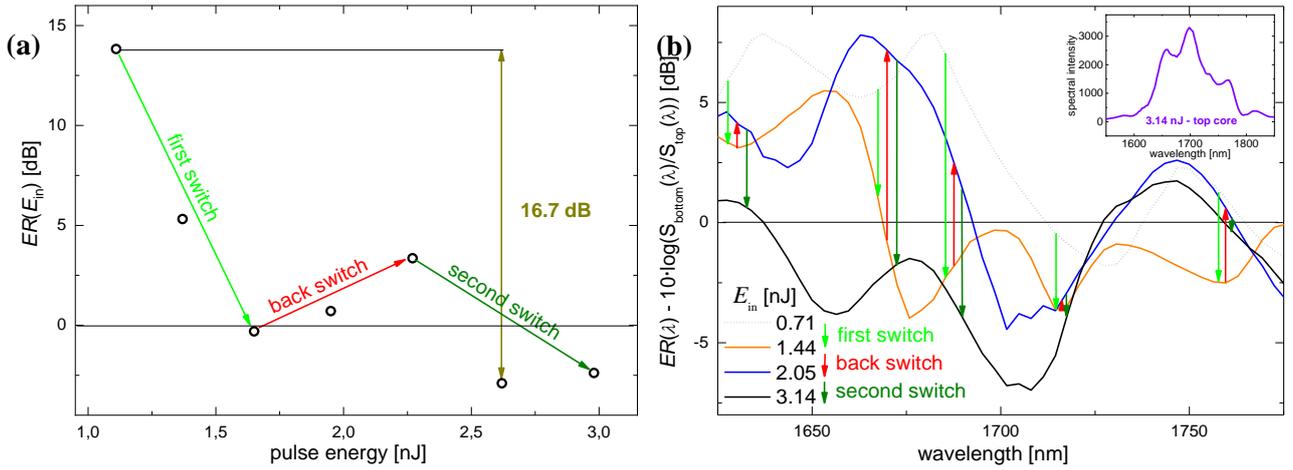

Fig.6 (a) Dual-core extinction ratio dependence on the input pulse energy acquired by processing the camera data in Fig.5. (b) Curves of spectrally resolved extinction ratio $ER(\lambda)$ at selected pulse energies. $S_{bottom}(\lambda)$ and $S_{top}(\lambda)$ were registered separately for both cores under identical experimental conditions as for the camera images. The three arrows (green, red and olive) denote respectively the first switch, back switch and second switch steps between the corresponding curves in different spectral areas. The spectrum recorded in the case of 3.14 nJ pulse energy and collected from the top core is shown in the inset.

Moreover, a double switching behaviour is apparent in Fig.6b. Here are also presented the $ER(\lambda)$ curves at intermediate pulse energies of 1.44 nJ and 2.05 nJ. The first switch, back switch and the second switch steps are marked with arrows between the corresponding curves. They are reported in consecutive order and directed always towards the increasing $E_{in}$. The first switching in the energy range 0.71 nJ – 1.44 nJ is rather convincing: the higher energy curve is situated below the low energy one and mostly in the negative extinction ratio section. The back-switching step in the energy range 1.44 – 2.05 nJ has also proper character, with some exception around 1700 nm and at the low intensity wings. Even in the case of the second switching (energy range of 2.05 – 3.14 nJ), the higher energy curve is nearly completely located below the lower energy one, with a major displacement in the negative and positive sections, respectively. Thus, the spectral profiles of the extinction ratio also confirm the double switching behaviour, which initially was observed by camera registrations. The results from the measured spectra exhibit lower global switching contrast values than the camera data, which was at the level of 16.7 dB. The spectrally resolved $ER$ difference between the two black curves (0.71, 3.14 nJ) is at maximum at the level of 10 dB. However, it is obvious that the spectral measurement is more distorted because of the chromatic aberrations affecting differently the $S_{bottom}(\lambda)$ and $S_{top}(\lambda)$ during their separate registration. Moreover, the spectra were affected also by power fluctuations of the OPA between the two consecutively registered energy-dependent spectral series. These were acquired separately for the two output chanells after the realignment of the whole registration part of the apparatus. Therefore, the IR camera results, which are free from the above-mentioned distortions, are more appropriate for the evaluation of the integral switching contrast. One can certainly expect even more convincing spectral results and higher values of spectrally resolved switching contrast in the case of distortion-free experimental conditions. Despite of the described drawbacks, the spectral results convincingly revealed the broadband character of the observed novel switching performance. It is also worth mentioning that the effect of the fibre losses is negligible taking into consideration of the measured average value for losses of 0.008 ± 0.001 dB/mm and the 35 mm fibre length.

From a general point of view, the observed character of the nonlinear switching has several advantages in comparison to the previous works. First, the value of the integral energy switching contrast of 16.7 dB is the highest observed so far in the case of DCF based nonlinear couplers [1,10]. Second, the spectral registrations showed that the switching performance has a broadband character. Finally, this is the first demonstration of double switching behaviour in the case of DCFs, simultaneously by camera and spectral registration. In the past, significant switching contrasts were observed only in limited areas of the nonlinearly broadened spectra [11,16]. Therefore, the results presented here have a key importance from the application point of view. These indicate the solitonic propagation regime effect, in which case the theoretical works predict nonlinear spatial transformations of input pulses without their break-up. However, up to now this broadband switching character has been never demonstrated in the case of DCF-based couplers in the anomalous spectral region. Based on the obtained results, the solitonic propagation concept is supported by the following findings: a) the excitation wavelength was in the middle of the DCF anomalous dispersion region; b) the pulse energies was rather moderate, around 1 nJ taking into consideration the 50 % in-coupling efficiency; c) the double switching exhibited broadband character in its all three steps; d) only moderate spectral broadening was registered with rather smooth spectra even in the case of highest applied pulse energy. All these observations indicate, that low order solitonic pulses evolved at the studied experimental conditions, which enabled their redirection forth and back between the two channels. Of course, the application potential and the solitonic regime should be confirmed by separate time-domain diagnostics of the two output channels. From the recent spectral registrations, it is obvious, that both the power stability of the source and the registration optical scheme should be improved to achieve this goal. The power stability quest is manageable if one uses oscillator based source instead of the presented complex amplifier – OPA system. However at 1700 nm wavelength there are no available ultrafast oscillators, therefore it is necessary to change the fibre design in order to establish the demonstrated propagation regime in C-band. Most importantly, the dual-core asymmetry should be improved, whose negative influence on the switching performance is enhanced at shorter wavelengths [16].

However we can state already, that the observed energy dependence character differs both from the previous experimental works and from the classical nonlinear directional coupler theory. This theory cannot explain high contrast broadband double switching under monotonic increase of pulse energy because it is based only on single switching by a nonlinear distortion of the dual-core symmetry [24]. Such process should be most effective exactly at $L_c$, which is 5 mm at 1700 nm for the studied fibre. However, our current experimental results reveal rich switching performance at 35 mm fibre length, which is 7 times longer than the $L_c$. Another interesting effect is that the double switching behaviour disappeared sequentially for both shorter and longer fibres. In fact, we haven't registered any switching effect at lengths of 25 mm or 45 mm. One possible interpretation of the presented observations is the switchable self-trapping of high order solitons, which promises broadband character and high levels of switching contrast [15,25]. In the frame of numerical studies of similar hexagonal core DCF, we obtained comparable optimal fibre length for high contrast switching at 32 mm considering 1700 nm excitation wavelength [19]. However there is some discrepancy between the numerical predictions and the experimental results regarding the switching energies. We demonstrated the first switching at significantly higher pulse energies in contrast to the predicted sub-nJ range, which took place in regime of self-trapped solitonic propagation. Another open question is the role of the dual-core asymmetry, which is always present in the case of real DCF couplers. Its effect was not addressed in the above mentioned theoretical work [21]. Combining more theoretical concepts, it is possible, that the first step of the double switching is based on the nonlinear elimination of the dual-core asymmetry [16,26] and the second step on the soliton self-trapping [15,25]. This concept is supported by our experimental observations in the case of bottom core excitation and identical experimental conditions. In that case, only single switching behaviour was observed. Therefore, we have the motivation to launch numerical simulations based on the real dual-core fibre structure, which will inherently incorporate the asymmetry effect as well. However, such complex study is already beyond the scope of the above experimental work and requires further inputs.

**Summary**

In this article, we demonstrated experimentally an efficient nonlinear self-switching of femtosecond pulses in special dual-core fibre in solitonic propagation regime. The specialty of the fibre is in the combination of two intentionally developed, thermally matched soft glasses with index contrast at the level of 0.4. We optimized the fibre structure for the envisaged switching performance in the low pulse energy range. Its microdimension cores were made of highly nonlinear glass. In the frame of complex cut-back study, the experimental investigation brought novel switching performance at optimal 35 mm fibre length using 1700 nm, 100 fs excitation pulses. Both camera and spectral registrations of the two output channels confirmed double switching behaviour. The global switching contrast was at the level of 16.7 dB. It was calculated from camera extinction ratios for pump pulse energies of 0.89 nJ and 2.62 nJ. Moreover, the processed spectral curves revealed a broadband character of the switching, with promising high application potential in the field of all-optical signal processing. The obtained results represent a significant progress in comparison to previous experimental works, as they reveal high switching contrast, a broadband switching character and a double switching behaviour. These achievements indicate the first realisation of a special switching principle based on the soliton self-trapping, which was predicted by theoretical works in similar conditions. We are planning further systematic investigation of this novel phenomenon in the future, both by dedicated numerical simulations and by time-domain study of the output field at improved experimental conditions.

**Acknowledgement**

This work was supported by National Science Centre, Poland [project No. 2016/23/P/ST7/02233 under POLONEZ program which has received funding from the European Union's Horizon 2020 research and innovation program under the Marie Skłodowska-Curie grant agreement No 665778], by Slovak R&D

Agency under the contracts No. APVV-17-0662 and SK-AT-2017-0026, by Austrian Agency OeAD under contract number SK 02_2018, by internal grant of Łukasiewicz Research Network - Institute of Electronic Materials Technology with No. S5-10-1024-19 and by EU under program H2020 ACTPHAST 4.0 (Grant No. 779472). I.A. A.P. and A.B. acknowledge financial support of the Austrian Research Promotion Agency FFG (Eurostars Eureka Project Nr 12576 HABRIA; FFG Project Nr 867822) and Austrian Science Fund FWF (P 27577-N27).